\documentclass[aps,prd,twocolumn,groupedaddress,nofootinbib]{revtex4-1}

\usepackage{graphicx,color,amsmath}
\usepackage{amsfonts}
\usepackage[export]{adjustbox}
\usepackage{hyperref}
\usepackage{mathtools} 

\definecolor{c1}{rgb}{0.121569, 0.466667, 0.705882}
\definecolor{c2}{rgb}{1., 0.498039, 0.054902}
\definecolor{c3}{rgb}{0.172549, 0.627451, 0.172549} 
\definecolor{c4}{rgb}{0.839216, 0.152941, 0.156863}
\definecolor{c5}{rgb}{0.580392, 0.403922, 0.741176}
\definecolor{c6}{rgb}{0.54902, 0.337255, 0.294118} 
\definecolor{c7}{rgb}{0.890196, 0.466667, 0.760784}
\definecolor{c8}{rgb}{0.498039, 0.498039, 0.498039}
\definecolor{c9}{rgb}{0.737255, 0.741176, 0.133333}
\definecolor{c10}{rgb}{0.0901961, 0.745098, 0.811765}

\begin{document}

\title{Cosmological Tension of Ultralight Axion Dark Matter and its Solutions}

\author{Jeff A. Dror}
\email{jdror@berkeley.edu} 
\affiliation{Theory Group, Lawrence Berkeley National Laboratory, Berkeley, CA 94720, USA}
\affiliation{Berkeley Center for Theoretical Physics, University of California, Berkeley, CA 94720, USA}

\author{Jacob M. Leedom}
\email{leedoj@berkeley.edu} 
\affiliation{Theory Group, Lawrence Berkeley National Laboratory, Berkeley, CA 94720, USA}
\affiliation{Berkeley Center for Theoretical Physics, University of California, Berkeley, CA 94720, USA}

\date{\today}

\begin{abstract}
A number of proposed and ongoing experiments search for axion dark matter with a mass nearing the limit set by small scale structure (${\cal O} ( 10 ^{ - 21 } {\rm eV} ) $). We consider the late universe cosmology of these models, showing that 
requiring the axion to have a matter-power spectrum that matches that of cold dark matter constrains the magnitude of the axion couplings to the visible sector. Comparing these limits to current and future experimental efforts, we find that many searches require axions with an abnormally large coupling to Standard Model fields, independently of how the axion was populated in the early universe. We survey mechanisms that can alleviate the bounds, namely, the introduction of large charges, various forms of kinetic mixing, a clockwork structure, and imposing a discrete symmetry. We provide an explicit model for each case and explore their phenomenology and viability to produce detectable ultralight axion dark matter.
 \end{abstract}

\maketitle

\section{Introduction}
Axions with masses well below the electroweak scale are simple dark matter candidates~\cite{Preskill:1982cy,Abbott:1982af,Dine:1982ah} with novel experimental signatures~\cite{Peebles:2000yy,Khmelnitsky:2013lxt,Hui:2016ltb}, a potential solution to the apparent incompatibility of cold dark matter with small scale structure~\cite{Hu:2000ke,Du:2016zcv}, and a common prediction of string theory~\cite{Arvanitaki:2009fg,Halverson:2019cmy}. Axions can arise as pseudo-Goldstone bosons of a spontaneously broken global symmetry or as zero modes of antisymmetric tensor fields after compactification of extra dimensions. In either case, their parametrically suppressed mass can result in a length scale comparable to the size of dwarf galaxies, the so-called {\em fuzzy} dark matter regime. Dark matter with a macroscopic Compton wavelength allows for novel detection opportunities and many on-going and future experimental efforts search for ultralight axion relics with masses around this limit. Purely gravitational searches look for erasure of structure on small scales as a consequence of the axion's sizable wavelength, limiting the axion mass, $ m _a \gtrsim {\cal O} ( 10 ^{ - 2 1} {\rm eV} )  $~\cite{Kobayashi:2017jcf,Irsic:2017yje,Nori:2018pka,Leong:2018opi,1808893,Schutz:2020jox,Nadler:2020prv}\footnote{Searches for oscillations in the local stress-energy tensor detectable with pulsar timing arrays~\cite{Khmelnitsky:2013lxt} set slightly weaker bounds~\cite{Porayko:2018sfa,Kato:2019bqz}.}. If axion dark matter has sizable non-gravitational coupling to the visible sector, there are additional detection strategies dependent on the nature of the coupling. 

Axion couplings to the Standard Model fields are intimately related to the shape of the axion potential. For a given axion decay constant, $ f _a $, axion couplings to matter are suppressed by $ \sim 1 / f _a $ while the ratio of the axion mass to its quartic coupling is typically at least on the order of $ \sim  f _a $. Such non-quadratic contributions to the axion potential play an important role in cosmology and the interplay between the axion's couplings and potential is the focus of this work. 

Measurements of the matter-power spectrum~\cite{Gil-Marin:2014sta,Aghanim:2018eyx,Abbott:2020knk} find that the cosmic energy density of dark matter redshifts as $ \propto R ^{-3} $, where $ R $ is the scale factor, until well before recombination. Such a scaling is satisfied for an oscillating scalar field if (and only if) the scalar potential is solely composed of a mass term, $ \frac{1}{2} m _a ^2 a ^2 $. This potential is typically only a good approximation if the field amplitude is sufficiently small and may not hold for ultralight axions in the early universe. Deviation from a simple quadratic term results in a perturbation spectrum that is no longer scale-invariant, constraining the axion potential. This in turn places a powerful bound on conventional ultralight axion dark matter candidates due to the relationship between $ f _a $ and the axion-matter couplings.

While this point is implicitly acknowledged in some of the axion literature, its significance is not widely emphasized, and its implications for current and future searches is missing altogether. In this work, we provide the constraints from the matter-power spectrum and thereby motivate a natural region of mass and coupling values wherein the axion can constitute dark matter without any additional model building. Circumventing the cosmological bounds requires breaking the parametric relationships between the axion potential and couplings. There are few techniques that can successfully accomplish this task. We consider the possibility of large charges, kinetic mixing, a clockwork structure, and discrete symmetries in the context of ultralight dark matter. These models typically predict additional light states in the spectrum and we survey their phenomenology. 

The outline of the paper is as follows. In section~\ref{sec:masscoupling} we review features of axion models, with particular emphasis on the coupling of axions to visible matter and the axion potential. In section~\ref{sec:power} we study the impact of ultralight axion dark matter on the matter-power spectrum and derive the associated bound. In section~\ref{sec:bounds} we examine the axion detection prospects of various experiments in light of the bounds.  In section~\ref{sec:robust} we study the robustness of the constraints by exploring ways to disrupt the relationship between axion-matter couplings and the axion potential. Finally, we conclude in section~\ref{sec:conc}.
\section{Axion mass and coupling}
\label{sec:masscoupling}
The axion decay constant, $f_a$, relates the terms of the axion potential to its couplings with Standard Model fields. The potential arises from non-perturbative contributions of gauge or string theories and explicitly breaks the continuous shift symmetry of the axion. We use the standard parametrization of the potential, which is a simple cosine of the form
\begin{equation} 
V  ( a ) \simeq  \mu  ^4  \cos \frac{ a }{  f _a }\,,
\label{eq:pot}
\end{equation} 
where $ \mu  $ is a scale associated with the explicit breaking of the global symmetry. If the potential arises from a composite sector (as in the case of the QCD axion), the explicit breaking scale corresponds to the maximum scale at which states must show up in the spectrum. The full axion potential is expected to be more complicated than the simple cosine above, but we can consider~\eqref{eq:pot} to be the first term in a Fourier decomposition of the potential. An important feature of ~\ref{eq:pot} is the existence of terms beyond the mass term and that the coefficients of these higher order terms are not arbitrary. The size of the quartic determines the point at which the quadratic approximation breaks down and has significance for axion cosmology.

Axions may also couple to Standard Model fields, with the leading operators obeying the axion shift symmetry. In this work we focus on two types of operators for axion dark matter, the prospective photon and nucleon couplings,\footnote{Other couplings of ultralight axions with matter are the gluon operator, $ a G _{\mu\nu} \tilde{G} ^{\mu\nu} $, the electron operator, $ \partial _\alpha  a \bar{e} \gamma ^\alpha  e $, and the muon operator, $ \partial _\alpha  a \bar{\mu} \gamma ^\alpha  \mu $. The gluon operator requires tuning to be sizable around the fuzzy dark matter regime and has other constraints~\cite{Hook:2017psm,Poddar:2020qft,Poddar:2019zoe}, the electron coupling can be probed using torsion pendulums~\cite{Graham:2017ivz}, and the muon operator has other strong constraints making it difficult to see experimentally \cite{Graham:2020kai,Janish:2020knz}.}

\begin{equation} 
{\cal L} \supset \frac{  C _{ a \gamma }\alpha }{ 8\pi f _a } a F \tilde{F}  + \frac{ C _{ a N} }{ f _a }\partial _\mu a \bar{N} \gamma ^\mu \gamma _5 N\,, 
\label{eq:acoupling}
\end{equation} 
where here and throughout we suppress the Lorentz indices on the gauge interactions, $ F \tilde{F} \equiv F _{\mu\nu} \tilde{F} ^{\mu\nu} $ and $ \tilde{F} ^{\mu\nu}  \equiv \frac{1}{2} \epsilon ^{ \mu \nu \alpha \beta } F _{ \alpha \beta } $. The parameters $ C _{ a \gamma } $ and $ C _{ a N } $ represent combinations of couplings in the UV theory and are $ {\cal O} ( 1 ) $ for generic axions. Demanding that the theory be invariant under axion discrete shift transformations requires the coefficient $C_{a\gamma}$ to be an integer~\footnote{If there is additional axion coupling in the phase of the mass matrix of some new fermions, $C_{a\gamma}$ only needs to sum to an integer with the coefficient of the coupling, see e.g.,~\cite{Fraser:2019ojt}} and hence cannot represent a large ratio of scales without additional model building. There may also be contributions to the above couplings from the IR if the axion mixes with dark sector particles, similar to the QCD axion-meson mixing, but such contributions will be unimportant for our considerations. The relationship between the axion potential and its coupling to matter is made manifest in~\eqref{eq:pot} and~\eqref{eq:acoupling}. To make contact with other studies, we define
\begin{equation} 
g_{a\gamma} \equiv \frac{ C_{a\gamma}\alpha }{  2\pi f_a} \quad , \quad g_{aN}\equiv \frac{C_{aN}}{f_a}\,.
\label{eq:can}
\end{equation} 

In principle it is possible that the particle searched for by dark matter experiments is not a true axion, in the sense that it is not shift symmetric, but a light pseudoscalar with a potential,
\begin{equation} 
{\cal L} _a = \frac{1}{2} m _a ^2 a ^2 \,.
\label{quad}
\end{equation} 
In this case the corresponding coefficients in-front of the terms in~\eqref{eq:acoupling} do not correspond to any symmetry breaking scale, but are instead completely free parameters associated with the scale of integrating out heavy fields. This would prevent us from using the arguments of section~\ref{sec:power} to restrict the dark matter parameter space. While such  models seem viable, they are highly fine-tuned and do not exhibit the desirable features of axion models. One way to see the tuning is to consider the additional terms in the effective theory that arise when integrating out the heavy fields that lead to the couplings in~\eqref{eq:acoupling}. For example, in addition to the $ a F \tilde{F} $ term, the low energy theory of a simple pseudoscalar will include terms such as $ a ^2 F F $, $ a ^3  F\tilde{F} $, etc. These terms will always be generated as they are no longer forbidden by any symmetry, and lead to corrections to the scalar potential which destabilize the light scalar. Thus any simple pseudoscalar becomes unnatural and the motivation to consider such a particle as dark matter is rendered null. Therefore, we take the position that the target particles of experimental searches are indeed ultralight axions with a full trigonometric potential, and we now examine the cosmological limitations of such dark matter candidates.

\section{Axion matter-power spectrum}
\label{sec:power}
A scalar field evolving in a purely quadratic potential has a scale-invariant matter-power spectrum, matching that of $ \Lambda {\rm CDM} $. However, if the potential contains higher order terms, the scalar equation of motion will possess non-linear terms which impact the growth of perturbations, with positive (negative) contributions wiping out (enhancing) small scale structure. For an axion with field amplitude $ a _0 ( z ) $ at redshift $ z $ the condition for the axion fluid to behave like cold dark matter is $ a _0 ( z ) / f _a \ll 1 $. The cosmic microwave background is the most sensitive probe of the matter-power spectrum, measuring deviations at a part per thousand, and sets a bound around recombination on any additional energy density fluctuations, $ \delta \rho / \rho \lesssim 10 ^{ - 3} $, corresponding to, $ a _0 ( z _{ \rm rec} ) ^2  / f _a ^2  \lesssim 10 ^{ - 3} $. It is important to note that this bound does not rely on the specific production mechanism and must be satisfied for any light axion making up the entirety of dark matter.

This constraint was studied quantitatively for misaligned axions in a trigonometric potential in~\cite{Poulin:2018dzj} (see also \cite{Desjacques:2017fmf} for related discussions). The authors considered an axion with the potential in~\eqref{eq:pot} and a field value frozen by Hubble friction until $ z _c $, the redshift at which the axion mass is comparable to the Hubble rate and oscillations begin. The matter-power spectrum then constrains the fraction of dark matter made up by axions as a function of $z_c$. The authors of~\cite{Poulin:2018dzj} find that in order for the axion to constitute all of dark matter, $z_c$ must be $\gtrsim 9 \times 10 ^4 $. This can be translated onto a constraint on $ f _a $ by noting that the axion field amplitude is fixed today by the measured dark matter energy density with, $\rho_{\rm DM} ( z )  = \frac{1}{2} m_a^2 a _0  ( z ) ^2$. Since the amplitude redshifts as $a _0 ( z ) \propto ( 1 + z ) ^{3/2}$, requiring the axion to oscillate before it exceeds its field range, $a _0  ( z_c ) \lesssim  f_a$, requires,
\begin{align} 
f_a & \gtrsim  \sqrt{\frac{ 2\rho_{\rm DM} ( z _0 ) }{ m_a^2}}(1+z_c)^{3/2} \,,\\ 
\text{or} \qquad f _a & \gtrsim 1.2 \times 10 ^{ 13} ~{\rm GeV} \,\left( \frac{ 10 ^{ - 2 0} {\rm eV} }{ m _a } \right) \,.\hspace{1cm}
\end{align} 
The rough expressions motivated above, $ a _0 ( z _{ {\rm rec}} ) ^2 / f _a ^2  \lesssim 10 ^{ - 3} $, gives a similar result. Note that while~\cite{Poulin:2018dzj}  assumed a misalignment mechanism, it is more general, and will apply (approximately) to any axion dark matter production mechanism as suggested by the rough estimate.

The constraint proposed in this work utilizes the matter-power spectrum and is distinct from the work of~\cite{Arias:2012az}, which presented a bound assuming the misalignment mechanism. The limit in ~\cite{Arias:2012az} is derived by noting that the maximum energy stored in the axion potential is $ \sim \mu ^4 $ and, assuming a simple cosmology from the start of oscillations to recombination, demanding that this be less than the dark matter energy density at $ z _c $: $ \rho _{ {\rm DM}} ( z _c ) \lesssim \mu ^4  $. This restricts, $ \mu ^4 \lesssim {\rm eV} ^4 (   z _c/z _{\rm eq}  ) ^3  $, or equivalently,
\begin{equation} 
f _a \gtrsim 10 ^{ 17 } ~{\rm GeV} \left( \frac{   10 ^{ -21} ~{\rm eV} }{m _a } \right) ^{1/4} \hspace{0.15cm} ( {\rm misalignment})\,.
\end{equation} 
The authors of~\cite{Arias:2012az} also consider temperature-dependent axion masses which relax the misalignment constraint. While both types of bounds in~\cite{Arias:2012az} are more stringent than the matter-power spectrum bound, they are also less robust since they rely on misalignment and on having a simple cosmology from $z_c$ to recombination.

The bound on $ f _a $ can be translated into a bound on the coupling to photons and nucleons using the relations in~\eqref{eq:acoupling} for given values of $ C _{ a \gamma } $ and $ C _{ a N } $. The results are displayed in Figs.~\ref{fig:pspace} and~\ref{fig:nspace} in black for several values of the coefficients. Since generic axions models predict $C_{a\gamma}, C_{aN} $ that are at most $ {\cal O} (  1 )  $, this is a powerful bound on the ultralight axion parameter space. Additional model building beyond the minimal scenario is required to access regions with larger coupling. For comparison, we include the regions constrained assuming misalignment as wavy gray lines for different values of $ C _{ a \gamma} $ and $ C _{ a N } $. 

The constraints we derive here assumed the axion field makes up dark matter until prior to recombination. An alternative scenario is the case where an axion is only produced at late times, such as through the decay of a heavier state. While an intriguing possibility, decay of heavy states will produce relativistic axions which will in turn modify the equation of state of the universe. Thus evading the matter-power spectrum bound by tweaking cosmology at late times is a formidable task. 

\section{Comparison with experiments}
\label{sec:bounds}
We now consider the prospects of ultralight axion dark matter searches in light of the matter-power spectrum restriction derived above, starting with a summary of current experimental constraints. Firstly, the Lyman-$ \alpha $-flux power spectra sets a bound on the axion mass, independent of the size of non-linear terms in the potential. These measurements are sensitive to sharp features in the matter-power spectrum on small scales, which would be present if the axion has a mass comparable to the size of dwarf galaxies, and set a bound on the axion mass of $ m _a \gtrsim 10 ^{ - 21} ~{\rm eV} $~\cite{Kobayashi:2017jcf,Irsic:2017yje,Nori:2018pka,Leong:2018opi}. A mass bound of similar magnitude can be determined by utilizing  constraints on the subhalo mass function from gravitational lensing and stellar streams~\cite{Schutz:2020jox,Nadler:2020prv}. Recently the Lyman-$\alpha$ bound was re-analyzed and strengthened to $m_a\gtrsim 2 \times 10^{-20}$ eV\cite{1808893}. Since the precise restriction on the axion mass varies between these studies, and so we include both the weakest and strongest bounds in the plots below. In addition, there are astrophysical bounds on axions that are independent of their energy density. Axions released during supernova (SN) 1987A would have produced a flux of axions that could convert as they passed through the galactic magnetic fields~\cite{Brockway:1996yr,Grifols:1996id,Payez:2014xsa} and the non-observation of this conversion sets the strongest bounds on low mass axions coupled to photons. For axion-nucleon coupling, the strongest dark matter-independent bounds arise from forbidding  excess cooling of SN-1987A~\cite{Chang:2018rso} and neutron stars~\cite{Sedrakian:2015krq,Hamaguchi:2018oqw,Beznogov:2018fda}. There are also the bounds arising from black hole superradiance~\cite{Arvanitaki:2010sy,Arvanitaki:2014wva}, but these are relevant for larger masses or smaller couplings than we consider.

There are a large number of searches looking for axion dark matter that rely on its relic abundance. Efforts to discover a photon coupling include looking for deviations in the polarization spectrum of the cosmic microwave background~\cite{Harari:1992ea} (with updated bounds in~\cite{Fedderke:2019ajk}), searching for the axion's influence on the polarization of light from astrophysical sources~\cite{Ivanov:2018byi,Fujita:2018zaj,Liu:2019brz,Caputo:2019tms}, and terrestrial experiments~\cite{Obata:2018vvr,Tobar:2020kmz,Berlin:2020vrk}.\footnote{We used the ``realistic'' projections of~\cite{Obata:2018vvr}.} Searches for an axion-nucleon coupling focused on the ultralight regime include axion-wind spin precision~\cite{Abel:2017rtm}, using nuclear magnetic resonance~\cite{Vasilakis:2008yn,Budker:2013hfa,Wu:2019exd,Garcon:2019inh,Bloch:2019lcy}, and using proton storage rings~\cite{Graham:2020kai}. Several spin precession experimental setups are considered in~\cite{Graham:2017ivz}.~\footnote{We selected the most stringent bounds from ~\cite{Graham:2017ivz,Graham:2020kai,Berlin:2020vrk} and continued these bounds to lower ULA mass values than what the original works consider.} 

\begin{figure*}[]
\centering
\includegraphics[width=0.75\textwidth]{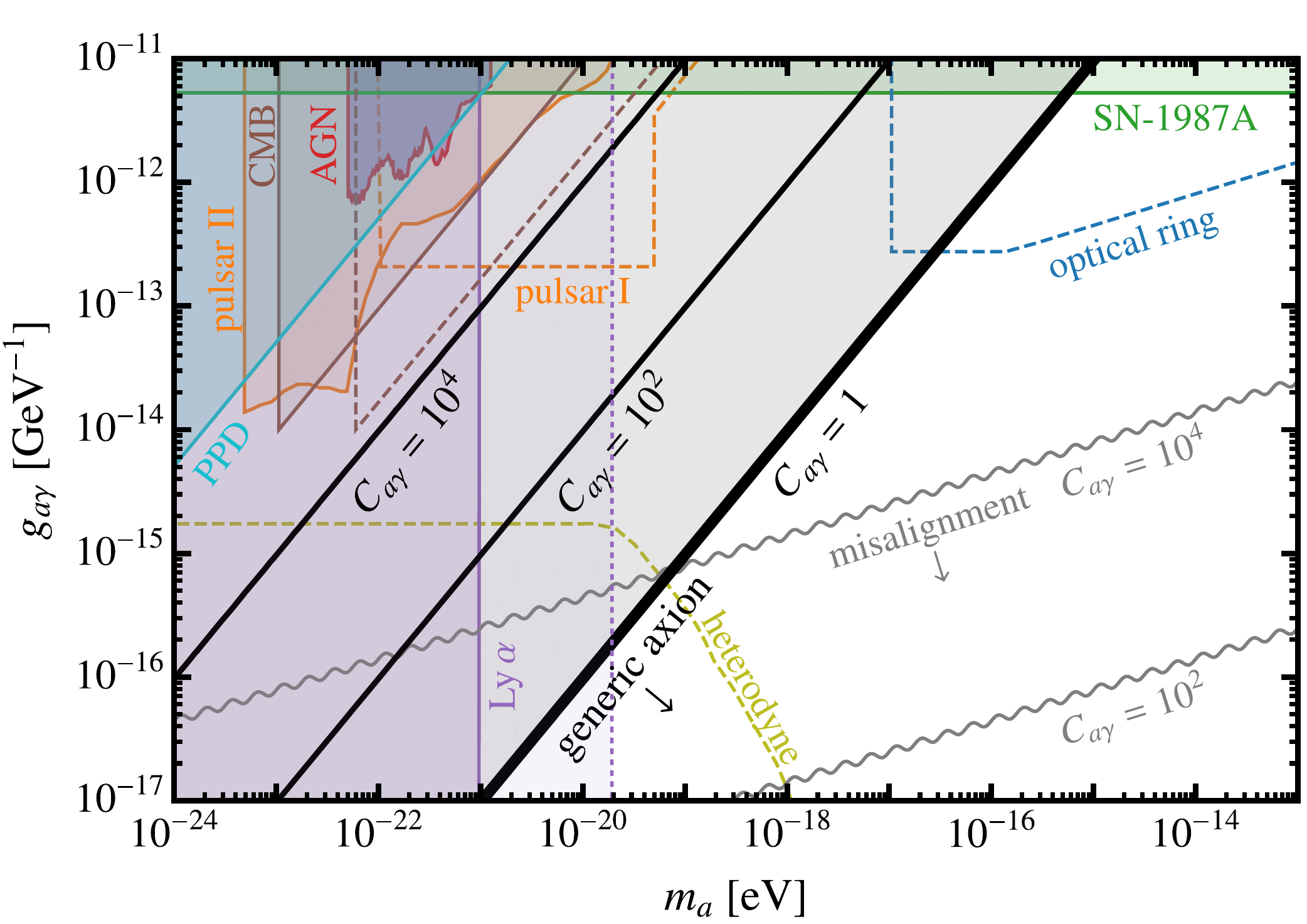}
\caption{Ultralight axion dark matter mass vs. photon-coupling parameter space. Requiring the ultralight axion to exhibit a matter-power spectrum consistent with $ \Lambda {\rm CDM} $ sets the bound shown in black for serveral values of $ C _{ a \gamma} $. The region below the $ C _{ a \gamma } = 1 $ line permits natural axions without any additional model building (see text).
The {\color{c5} {\bf solid purple}} and {\color{c5} {\bf dotted purple}} lines display the weakest and strongest bounds, respectively, arising from purely gravitational considerations ~\cite{Sedrakian:2015krq,Hamaguchi:2018oqw,Chang:2018rso,Beznogov:2018fda}.The lack of axion-to-photon conversion of axions produced during supernova-1987A~\cite{Payez:2014xsa} gives the bound in {\color{c3} {\bf green}}. Additional bounds from current (solid) and proposed searches (dashed) are from active galactic nuclei~\cite{Ivanov:2018byi}({\color{c4} {\bf red}}), protoplanetary disk polarimetry~\cite{Fujita:2018zaj} ({\color{c10} {\bf light blue}}), CMB birefringence~\cite{Fedderke:2019ajk} ({\color{c6} {\bf brown}}), pulsars~\cite{Liu:2019brz,Caputo:2019tms} ({\color{c2} {\bf orange}}), optical rings~\cite{Obata:2018vvr} ({\color{c1} {\bf dark blue}}), and heterodyne superconductors~\cite{Berlin:2020vrk} ({\color{c9} {\bf olive}}). The misalignment bounds for $C_{a\gamma} = 10^2,10^4$ are displayed by the wavy contours ({\color{c8} {\bf grey}}).}
\label{fig:pspace}
\end{figure*}

\begin{figure*}[]
\centering
\includegraphics[width=0.75\textwidth]{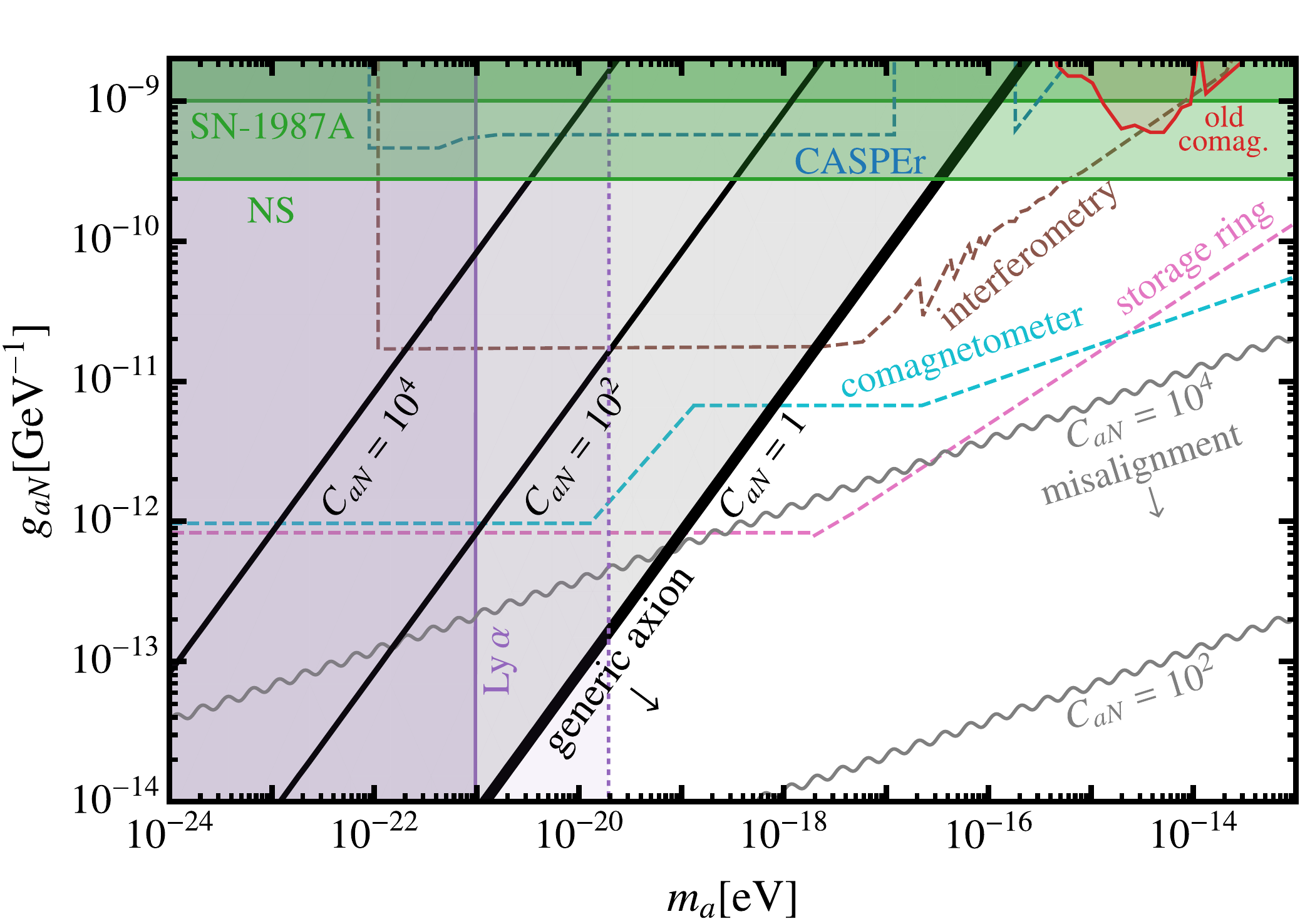}
\caption{Ultralight axion dark matter mass vs. nucleon-coupling parameter space. Requiring the ultralight axion to exhibit a matter-power spectrum consistent with $ \Lambda {\rm CDM} $ sets the bound shown in black for serveral values of $ C _{ a N} $. The region below the $ C _{ a N } = 1 $ line permits natural axions without any additional model building (see text). The {\color{c5} {\bf solid purple}} and {\color{c5} {\bf dotted purple}} lines display the weakest and strongest bounds, respectively, arising from purely gravitational considerations ~\cite{Sedrakian:2015krq,Hamaguchi:2018oqw,Chang:2018rso,Beznogov:2018fda}. Supernova-1987A \& neutron star cooling~\cite{Sedrakian:2015krq,Hamaguchi:2018oqw,Chang:2018rso,Beznogov:2018fda} bounds are shown in {\color{c3} {\bf green}} and the bound from old comagnetometer data is shown in {\color{c4} {\bf red}} \cite{Bloch:2019lcy}. Additional bounds for nucleon couplings are shown from projections of the CASPEr-Zulf experiment~\cite{Wu:2019exd,Garcon:2019inh} ({\color{c1} {\bf dark blue}}), atom interferometry ({\color{c6} {\bf brown}}) ~\cite{Graham:2017ivz}, atomic magnetometers ({\color{c10} {\bf light blue}})~\cite{Graham:2017ivz}, and storage rings~\cite{Graham:2020kai} ({\color{c7} {\bf pink}}). The misalignment bounds for $C_{a N} = 10^2,10^4$ are displayed by the wavy contours ({\color{c8} {\bf grey}}).}
\label{fig:nspace}
\end{figure*}

The photon bounds are compiled in Fig.~\ref{fig:pspace} and nucleon bounds in Fig.~\ref{fig:nspace}. The matter-power spectrum bound derived in section~\ref{sec:power} is displayed in both figures. We use solid (dashed) lines to denote current (prospective) bounds. We conclude that many experimental proposals in this ultralight regime are inconsistent with a generic axion dark matter and require $ C _{ a \gamma } \gg 1  $ or $ C _{ a N } \gg 1 $. Reaching the large couplings considered in various experiments is an issue of additional model building, and is the focus of the next section. 

\section{Enhanced axion couplings}
\label{sec:robust}
We have presented stringent bounds on axions arising from the relationship between their field range, $ f _a $, and their coupling to photons or nucleons. However, there exist model-building techniques that can relax this relationship, which have often been discussed in the context of axion inflation. These methods may also be applied to ultralight axion dark matter and have distinct low energy phenomenology as a consequence of the lightness of the axion and the requirement of matching the observed matter-power spectrum. In this section we review these mechanisms, provide explicit realizations of such models, and study their phenomenology. We focus on the photon coupling, but similar models can be built for the nucleon coupling. 

\subsection{Large Charges}
\label{subsec:charge}
One way to enhance the axion coupling to visible matter is to introduce fermions with large charges or a large number of fermions (see e.g., ~\cite{Agrawal:2017cmd,Agrawal:2018mkd} for a discussion in the context of inflation). This strategy is limited by the requirement of perturbativity of electromagnetism and the presence of light fermions charged under electromagnetism.

To be explicit, consider a KSVZ-like model~\cite{Kim:1979if,Shifman:1979if} where a complex scalar $ \Phi  $ (whose phase will be identified with the axion), has Yukawa couplings with a set of Weyl fermions with an electromagnetic charge $ Q _f   $. Integrating out the fermions leads to an axion-photon coupling,
\begin{equation} 
{\cal L} \supset \frac{ \alpha Q   ^2 _f N _f }{ 8\pi f _a } a F \tilde{F}.
\end{equation} 
The presence of charged fermions renormalizes the electric charge as computed through corrections to the photon gauge kinetic term. Perturbativity requires that $N_fQ^2 _f \alpha/4\pi \lesssim 1 $. Since $C_{a\gamma}  = N_f Q _f ^2 $, the perturbativity constraint sets a bound $C_{a\gamma} \lesssim 4\pi /\alpha$. We conclude that large charges can at most enhance the axion-photon coupling by $ {\cal O} ( 10 ^3 )  $.\footnote{ Here we have taken the Peccei-Quinn charges of the fermions to be $\mathcal{O}(1)$. If one chooses larger Peccei-Quinn charges such that the fermion mass only arises through higher dimensional operators, then the photon coupling can be slightly amplified. However, requiring a hierarchy between $f_a$ and the cutoff strongly constrains this possibility~\cite{Agrawal:2017cmd}.}

\subsection{Kinetic Mixing}
Kinetic mixing of multiple axion fields can raise the axion coupling to visible matter by (potentially) allowing an axion with a large field range to inherent couplings of an axion with a smaller field range (see~\cite{Babu:1994id,Bachlechner:2014hsa,Shiu:2015uva,Higaki:2014qua,Cicoli:2012sz,Shiu:2015xda,Agrawal:2017eqm,Agrawal:2017cmd,Agrawal:2018mkd} for discussions in other contexts). As a simple example consider two axions $a_1$ and $a_2$, where $a_1$ obtains a potential while the lighter axion, $a_2$ (which is massless here), couples to photons: 
\begin{align}
\mathcal{L} \supset \frac{1}{2}\partial_\mu a_1 \partial^\mu a_1 &+ \frac{1}{2}\partial_\mu a_2 \partial^\mu a_2  + \varepsilon\,\partial_\mu a_1\partial a_2\nonumber \\
&+ \mu^4 \cos\frac{a_1}{F_1} + \frac{\alpha}{8\pi F_2}a_2F_{\mu\nu}\tilde{F}^{\mu\nu}\,.
\label{eq:kineqn}
\end{align}
The kinetic term can be diagonalized by the shift $a_2\rightarrow a_2 -\varepsilon a_1$, which induces an $a_1$-photon coupling,
\begin{equation}
\mathcal{L} \supset -\frac{\varepsilon F_1}{F_2}\frac{\alpha}{8\pi F_1}a_1F_{\mu\nu}\tilde{F}^{\mu\nu}\,.
\end{equation}
Taking $a _1 $ to be the  axion dark matter candidate, we conclude that kinetic mixing gives $ C _{ a \gamma } = \varepsilon F_1/F_2$. If $ \varepsilon $ is held fixed and the decay constants have a large hierarchy ($ F _1 \gg F _2 $), then $ a  _1 $ will have $ C _{ a \gamma } \gg 1 $. 

While this appears to be a simple solution, it is not possible to have 
$C_{a\gamma}\gtrsim 1$ within most field theories. This is a consequence of axions arising as Goldstone bosons of a extended scalar sector and hence the axion kinetic mixing is not a free parameter but must be generated. There are two possible sources for $ \varepsilon $: renormalization group flow (``IR'') and higher dimensional operators (``UV'') contributions. To see the suppression from IR contributions, consider a theory of two axions with a fermion, $ \chi $,
\begin{equation}
\mathcal{L} \supset  \frac{\partial^\mu a_1}{F_1}\bar{\chi } \gamma_\mu \gamma_5 \chi  + \frac{\partial^\mu a_2}{F_2}\bar{\chi } \gamma_\mu \gamma_5 \chi \,.
\end{equation}
The induced kinetic mixing of the axion is quadratically divergent and goes as ,
\begin{equation} 
\varepsilon  \sim \frac{ \Lambda  ^2 }{  (4\pi)^2 F _1 F _2 }  \,,
\end{equation} 
where $ \Lambda $ represents the cutoff scale. Since $ \Lambda  \lesssim F _{ 1,2} $ (otherwise the effective theory is inconsistent), the kinetic mixing is bounded by $ \varepsilon \lesssim F _2 / 4\pi F _1 $ and will result in $ C _{ a \gamma } \lesssim 1 $. 

Alternatively, it is possible to induce an axion kinetic mixing through higher dimensional operators (see e.g., ~\cite{Babu:1994id,Higaki:2014qua}). Taking $a_1$ and $a_2$ to be the phases of complex scalar fields $\Phi_1$ and $\Phi_2$, there can be an operator,
\begin{equation}
\mathcal{L} \supset \frac{1}{2M ^2}\,\Phi_1^\dag \overset\leftrightarrow{ \partial } \Phi _1 \,\Phi_2^\dag \overset\leftrightarrow{ \partial } \Phi _2 \,,
\end{equation}
where $ \Phi ^\dagger  \overset\leftrightarrow{  \partial } \Phi \equiv \Phi ^\dagger  \partial  \Phi -  ( \partial\Phi ^\dagger  )     \Phi $. Once the scalar fields take on their vacuum values, the axions get a mixing term with $\varepsilon   = F_1F_2/M^2$. This is again suppressed since consistency of the effective theory requires $M \gtrsim F _{ 1,2} $ and cannot result in $ C _{ a \gamma } \gtrsim 1 $.

While these examples show that kinetic mixing is not typically sizable for field theory axions, it is has been suggested that certain string constructions allow for sizable mixing coefficients~\cite{Agrawal:2017cmd}. While we are not aware of a concrete string construction where this is true, this may be a way to achieve $ C _{ a \gamma } \gtrsim 1 $. 

Interestingly, for ultralight axion dark matter, kinetic mixing has additional phenomenological implications. In order for the Lagrangian in~\eqref{eq:kineqn} to result in a photon coupling for $ a _1 $ that is not suppressed by a ratio of axion masses, $ a _2 $ must be lighter than $ a _1 $. Since the $ a _2 $ photon coupling is not suppressed by factors of $ \varepsilon $, it may be more detectable than $ a _1 $ and drastically influence direct constraints, such as from supernova axion cooling or conversion. This would need to be studied with care for a particular realization of a value of $ \varepsilon $.

In addition to axion-mixing, kinetic mixing of abelian gauge fields can boost the axion-photon coupling, as considered in~\cite{Daido:2018dmu}. In this case the coupling may be enhanced if the axion-photon coupling inherits the dark photon gauge coupling. To see this explicitly, we consider an axion coupled to a dark U(1) gauge field, $A^{\prime}$, which kinetically mixes with the electromagnetism,
\begin{equation}
	\mathcal{L}\supset \frac{ \alpha ' }{ 8\pi F _a } a F ' \tilde{F}  '  -\frac{1}{4}F F -\frac{1}{4}F^\prime F^{\prime}- \frac{\epsilon }{4}F F^\prime,
\end{equation}
where $ \alpha ' $ is the dark gauge coupling. If $ A' $ has a mass below the photon plasma mass, then a basis rotation can be performed to diagonalize the kinetic terms through $ A \rightarrow A - \epsilon A' $. This transformation leaves the dark photon approximately massless and gives the axion a coupling to photons as
\begin{equation} 
	\mathcal{L}\supset \frac{ \epsilon ^2 \alpha '   }{ 8\pi F _a } a F  \tilde{F}  \,,
\end{equation} 
such that $ C _{ a \gamma } = \epsilon ^2 \alpha '  / \alpha  $. Direct constraints on dark photons permit $ \epsilon \sim 1 $ (see ~\cite{Mirizzi:2009iz,Caputo:2020bdy,Garcia:2020qrp} for the bounds on ultralight dark photons) while $ \alpha ' $ can be $ \sim 1 $. Taken together, gauge kinetic mixing permits an amplification factor $ C _{ a \gamma } \sim \mathcal{O}(10 ^2) $ .
 
So far we have considered the cases of axion-axion and vector-vector mixing. It is also possible for axions to mix with a vector if the axions transform under the gauge symmetry, as is the case for St\"{u}ckelberg axions (see, e.g., ~\cite{Shiu:2015xda,Shiu:2015uva} for discussions in the context of inflationary model building, as well as~\cite{Fraser:2019ojt,Choi:2019ahy}). As a simple model, we consider the case of two St\"{u}ckelberg axions that have gauge interactions with a dark U(1) gauge field, $ A ' $, and nearly identical interactions with electromagnetism and a dark confining gauge sector: 
\begin{align}
\mathcal{L} &\supset \frac{1}{2}\bigg(\partial_\mu a_1  - q_1F_1A_\mu'\bigg)^2 + \frac{1}{2}\bigg( \partial_\mu a_2  - q_2F_2A_\mu'\bigg)^2\nonumber\\
						&+ \frac{\beta \alpha_s}{8\pi}\bigg(\frac{a _1 }{F  _1 }  +\frac{ a_2 }{ F _2 } \bigg)G\tilde{G} + \frac{\alpha}{8\pi }\bigg(\frac{a _1 }{ F _1 }+\frac{a_2 }{ F _2 }\bigg)F\tilde{F}.
\label{stuc1}
\end{align}
Gauge invariance requires $q_2=-q_1\equiv-q$.\footnote{This Lagrangian is invariant under the U(1) gauge transformation $a_1\rightarrow a_1 +q_1F_1\alpha$, $a_2\rightarrow a_2+q_2F_2\alpha$, and $A_\mu\rightarrow A_\mu + \partial_\mu \alpha$ if $q_1+q_2=0$. The Lagrangian we consider is a simplified version of the setups in~\cite{Shiu:2015xda,Shiu:2015uva,Fraser:2019ojt}, but the conclusions are unchanged in the more general scenarios.} We perform a field redefinition,

\begin{align}
	a &= -\bar{F}\bigg(	\frac{a_1}{F_1}+\frac{a_2}{F_2}		\bigg)	\nonumber\\
b &= \bar{F}\bigg(\frac{a_1}{F_2}-\frac{a_2}{F1}\bigg),
\end{align}
so that the physical axion interactions are,
\begin{align}
\mathcal{L} \supset- \frac{\beta }{8\pi \bar{F}}aG\tilde{G} - \frac{1}{8\pi \bar{F}}aF\tilde{F},
\end{align}
and $\bar{F} = F_1F_2/(F_1^2+F_2^2)^{1/2}$. The axion $b$ remains charged and provides a mass for the dark gauge boson. The surviving axion, $a$, is neutral under the dark U(1) and is the dark matter candidate. Since $\bar{F} $ is smaller than $ F_1 $ and $ F _2 $, $a$ is more strongly coupled to photons than either of the original axions~\cite{Fraser:2019ojt}. Nevertheless, this does not result in $ C _{ a \gamma } \gtrsim 1 $. This is because the decay constant of the surviving axion, $\bar{F}$, appears in the anomalous coupling to both the non-abelian and electromagnetic gauge sectors and so the canonical relationship between axion potential and matter coupling is maintained. We conclude that axion-vector mixing cannot be used to evade the cosmological bounds on ultralight dark matter axions.

\subsection{Clockwork}
Clockwork models provide a means to disturb the canonical relationship between the axion potential and photon coupling by introducing a large number of axions, each interacting with both its own confining gauge sector and its ``neighbor''. After a rotation to the axion mass basis, the lightest axion's potential can be exponentially suppressed without introducing an exponential number of fields. This light axion can be understood as the Goldstone boson of a global symmetry between scalar fields in a UV completion (see, e.g.,~\cite{Choi:2015fiu,Farina:2016tgd,Coy:2017yex,Agrawal:2017cmd,Agrawal:2018mkd,Marques-Tavares:2018cwm} for discussions in different contexts). 

As an explicit model, we consider a set of $N$ axions, $a_i$, with couplings to $N$ SU($n_i$) gauge sectors with field strengths $G_i$, and a photon coupling only for $a_N$:
\begin{align}
\mathcal{L} \supset \sum_{i = 1 }^{N-1}\frac{\alpha_{s,i+1}}{8\pi}&\bigg(\frac{\beta _ia_i}{F_i}+\frac{a_{i+1}}{F_{i+1}}\bigg)G_{(i+1)}\tilde{G}_{(i+1)}\nonumber\\
					& +\frac{\alpha_{s,1}}{8\pi F_1}a_1G_1\tilde{G}_1+ \frac{\alpha}{8\pi F_N}a_N F \tilde{F}\,.
\end{align}
The $\beta _i$ factors are integers greater than or equal to unity and we have omitted a bare $\theta$ term. Upon confinement, the gauge sectors give rise to the potential for the axions,
\begin{equation}
V ( a _i ) \simeq  \sum_{i=1}^{N-1}\mu_{i+1}^4\cos\bigg(\frac{\beta _ia_i}{F_i}+\frac{a_{i+1}}{F_{i+1}}\bigg) +\mu_1^4 \cos\frac{a_1}{F_1} \,,
\end{equation}
where the $ \mu _i $ are the confinement scales and represent the maximum possible masses for the dark composite states.
To get an enhanced photon coupling, we require
\begin{equation} 
 \mu _{1}  \ll \mu _2 , \mu _3 , ... 
\end{equation} 
and we take the $ F _i $'s to be comparable to each other. In this case, up to $ {\cal O} ( \mu _1 / \mu _i )  $ corrections, integrating out the heavy axions corresponds to iteratively introducing the substitution:
\begin{equation} 
  \frac{\beta _ia_i}{F_i}+\frac{a_{i+1}}{F_{i+1}} \simeq 0 \quad \forall i = 1, 2, ... \hspace{.1cm}.
\end{equation} 
This transformation produces the effective Lagrangian
\begin{equation}
{\cal L}  \simeq  \mu_1^4 \cos \frac{a_N}{  F_N \prod _i  \beta _i  }   + \frac{\alpha}{8\pi F_N } a _N F \tilde{F}
\,.
\end{equation}
The $ a _N $-potential is exponentially suppressed by $ \prod _i \beta _i $ while the photon coupling remains unchanged, resulting in an axion potential exponentially flatter than the naive estimate. Redefining the axion decay constant as in \eqref{eq:pot} gives,
\begin{equation} 
C _{ a \gamma } = \prod _i  \beta _i  \,,
\end{equation} 
thereby boosting the photon coupling relative to a generic axion. 
 
We now consider the phenomenological implications of clockworked axions as dark matter. Firstly, in addition to the light axion, there exist $ N - 1  $ axions with masses proportional to $ \mu _{ i \ge 2 } $ (the only bound on these is from unitarity, requiring $ \mu _{ i } \lesssim F _i $\cite{Agrawal:2018mkd}). These would be populated in the early universe if the new non-abelian gauge groups confine after reheating (from their own misalignment mechanisms) or if they are thermalized. Assuming the confinement scales $ \mu _{ i \ge 2 } $ are comparable, the energy density of the heaviest axion would dominate.
However, the photon couplings of the $ N - 1 $ heavy axions are suppressed by products of $ \beta _i $ relative to the coupling of the lightest axion, and so they cannot be the target particles of the above experimental searches. Furthermore, if the lightest clockwork axion is to be dark matter and the experimental target, the heavier axions must decay into into Standard Model particles (if the axions decayed into lighter axions, they would produce excess dark radiation in conflict with measurements of $ \Delta N _{ {\rm eff}} $). This mandates the need for substantial couplings of the heavy axions to the Standard Model and may lead to observable effects in terrestrial experiments. 

In addition to the heavy axions, the clockwork model predicts the existence of a non-abelian gauge sector with composite states well below the electroweak scale and masses below $ \mu _1 $. Demanding that $F _N  <M_{\rm pl}$ results in a confinement scale of $ \sim 10~{\rm keV} \prod _i \beta _i $ for ultralight axion dark matter with a mass $ \mathcal{O}(10 ^{ - 20}) $ eV. Depending on the type of interactions this light gauge sector has with the Standard Model, it may be possible to observe these in terrestrial experiments. 

	From the low energy perspective, the clockwork model we have described appears to permit arbitrarily large $C_{a\gamma}$ values. However, there may be limitations on this enhancement factor if one attempts to embed the model into a string construction. In heterotic string models, the 4 dimensional gauge groups descend from the rank 16 gauge groups $E_8 \times E_8$ or $SO(32)$. Demanding that the Standard Model's rank 4 gauge group be present in the low energy theory restricts the rank of the dark sector to be $\leq 12$ and so $N$  could be severely limited~\cite{Ibanez:2017vfl}\footnote{We assume a generic Calabi-Yau compactification manifold.}.  We leave an extensive study of string compactification restrictions on clockwork models to future work.

\subsection{Discrete Symmetry}
Finally, axion couplings to visible matter can be augmented by introducing multiple non-abelian gauge sectors related by a discrete symmetry~\cite{Hook:2018jle}. When the axion potential contributions from the confinement of each gauge sector are summed together, one finds the potential may be exponentially suppressed compared to the naive expectation. 

As an example, we consider a theory with a single axion, $a$, that couples to $N$ confining gauge sectors with field strengths, $ G_{(n)} $, and impose a discrete symmetry under which,
\begin{align} 
 a & \rightarrow a +  2\pi F _a / N \notag \\ 
\hspace{.5cm} G _{ ( n ) } & \rightarrow G _{ ( n + 1 ) }\hspace{1.cm}\,.
\end{align} 
The symmetry forces all the non-abelian gauge sectors to share a common gauge coupling and fermion content. Including an axion-photon coupling, the Lagrangian consistent with the symmetry has,
\begin{equation}
\mathcal{L} \supset \frac{\beta \alpha_s}{8\pi}\sum_{n = 1} ^N  \bigg(\frac{a}{F_a}+\frac{2\pi n}{N}\bigg)G_{(n)}\tilde{G}_{(n)}+ \frac{  \alpha }{ 8\pi F _a } a F \tilde{F} \,.
\end{equation}
In contrast to clockwork, the integer $ \beta $ serves no essential purpose here and can be taken to be unity.

Each of the $N$ gauge sectors contribute to the axion potential after they confine. If we were to use the leading contribution to the axion potential from~\eqref{eq:pot} for each sector, the total axion potential would vanish. Therefore we must include corrections associated with higher modes in the Fourier expansion of the potential, which depend on the light fermion content of the theory. For a sector with two fermions with masses $ m _1 $ and $ m _2 $ below the composite scale, chiral perturbation theory yields the leading order potential (see, e.g.,~\cite{diCortona:2015ldu}), 
\begin{equation}
		V (a) = - \mu^4 \sum_{n=0}^{N-1}\sqrt{1-z\sin^2\bigg(\frac{a}{2F_a}+\frac{\pi n}{N}\bigg)}
\label{eq:dispot}
\end{equation}
with $z=4m_1m_2/(m_1+m_2)^2$. After the sum in \eqref{eq:dispot} is carried out, one finds the axion mass is exponentially suppressed if there is a small hierarchy between the light quark masses. Taking $ m _2 > m _1 $ the axion mass dependence on $N$ is, approximately, 
\begin{equation} 
m _a \sim \left( \frac{ m _1 }{ m _2} \right)  ^{   N/2 } \frac{ \mu ^2 }{ F _a } \,.
\end{equation} 
Canonically normalizing the decay constant, we get $ C _{ a \gamma } \sim ( m _2/ m _1 )  ^{N/2} $, breaking the relation between the axion mass and photon coupling for $ N \gg 1 $.

While discrete symmetries produce axions with $ C _{ a \gamma } \gg 1 $, they do not evade the bounds from the matter-power spectrum. This is a consequence of the axion potential from \eqref{eq:dispot} giving unusually large higher order axion terms. Unlike clockwork, which keeps the axion potential of the form in \eqref{eq:pot} and simply extends the field range, discrete symmetries break this relationship entirely. To see this behavior, we expand~\eqref{eq:dispot} about one of its minima, giving the potential,

\begin{align}
	V (a) &=  \frac{C_2}{2}\frac{\mu^4}{F_a^2}a^2 - \frac{C_4}{4!}\frac{\mu^4}{F_a^4}a^4+\cdots\,,\nonumber\\
			&= \frac{1}{2}m_a^2a^2 - \frac{1}{4!}\lambda a^4 +\cdots\,,
\end{align}
where the $C_i$'s are constants that arise from the sum in ~\eqref{eq:dispot}. The coefficient $C_2$ determines the exponential suppression of the axion mass and $C_4$ fulfills a similar role for the quartic. It is convenient to recast the mass suppression factor into an axion-photon coupling enhancement factor via $f_a \equiv  F_a/\sqrt{C_2}$ such that $m_a = \mu^2/f_a$,  $\lambda = C_4 \mu^4/ C _2 f_a^4$, and $C_{a\gamma} = \sqrt{C_2}$.

\begin{figure}[t]
\centering
\includegraphics[width=0.48\textwidth]{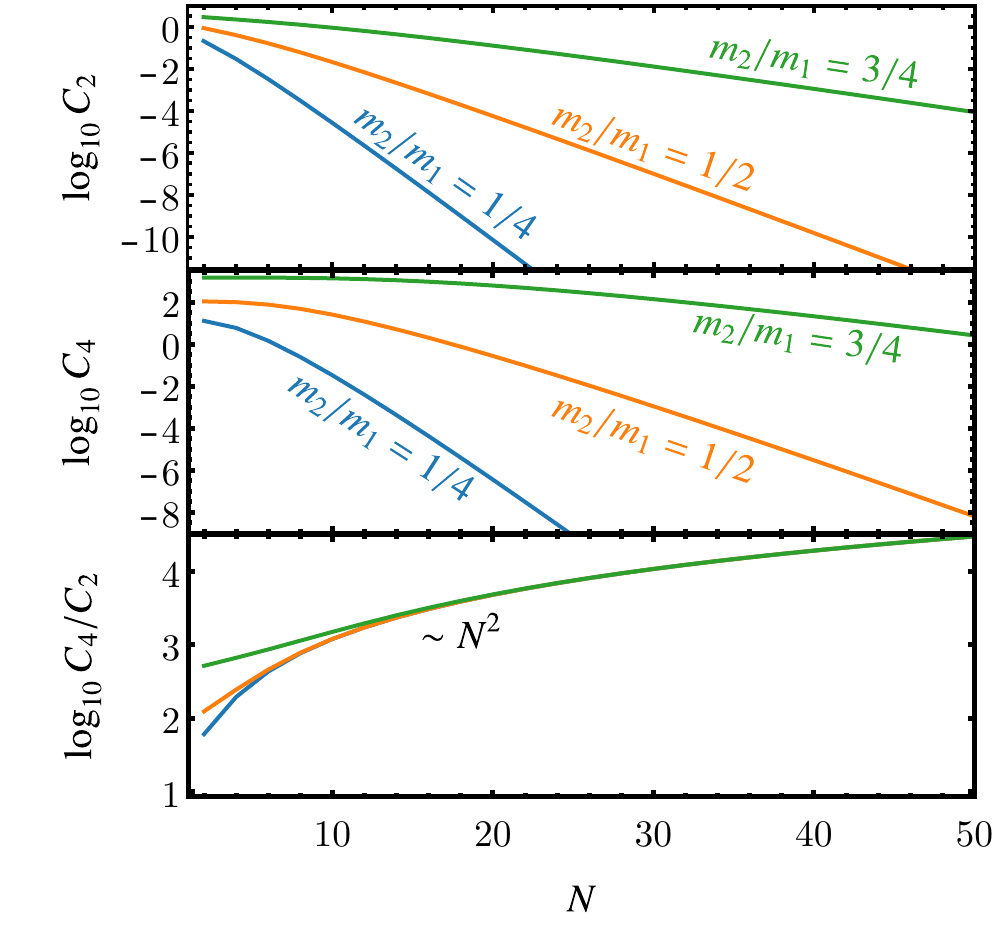}
\caption{The coefficients of the axion potential arising from a discrete symmetry normalized to their expected values. We see the exponential drop in $ C _4 $ and $ C _2 $, however their ratio grows with $ N $. Thus discrete symmetries strengthen the matter-spectrum bounds instead of weakening them  (see text).}
\label{fig:disco}
\end{figure}

The key observation is that the dependence on $N$ is different for the two constants $C_2$ and $C_4$, as displayed in Fig.~\ref{fig:disco}. For large $N$, $C_4$ decreases more slowly than $C_2$ with increasing $N$. The approximate condition presented above for the axion to behave sufficiently like cold dark matter is,

\begin{align}
			&\frac{\lambda a^4_0}{m_a^2a^2}\bigg\rvert_{\rm eq} \sim  \frac{\lambda \text{eV}^4}{m_a^4} = \frac{C_4}{C_2}\frac{\text{eV}^4}{m_a^2}\frac{C_{a\gamma} ^2 }{f_a ^2 }\lesssim 10^{-3}\,.
\end{align}
The factor $\text{eV}^4 C _{ a \gamma } ^2 /m_a^2 f _a ^2 $ is restricted to be greater than unity to get a large enhancement in the photon coupling. From Fig.~\ref{fig:disco}, we see that $C_4/C_2$ will also be greater than unity and the bound cannot be satisfied. We conclude that this variety of model cannot be used to boost the axion-photon coupling for ultralight axion dark matter. 

\section{Conclusions}
\label{sec:conc}
In this work, we considered the experimental prospects of detecting ultralight axion dark matter through its couplings to the visible sector, focusing on photon and nucleon interactions.  We presented a stringent bound on axions by requiring that their matter-power spectrum match that of $ \Lambda {\rm CDM} $ and concluded that generic axions are constrained to have couplings significantly smaller than is often assumed. This bound makes use of the relationship between axion-matter couplings and the axion potential and is independent of the dark matter production mechanism. This discussion displays the tension between experimental projections and cosmological bounds and has not been widely emphasized in previous literature. 

Given the up and coming experimental program, the need to understand the landscape of ultralight axion dark matter models with detectable couplings is clear. As such, we studied various strategies to boost axion couplings that were introduced previously in the literature and applied them to ultralight axion dark matter. In particular, we considered models with large charges, diverse forms of kinetic mixing, a clockwork mechanism, and a discrete symmetry. We examined the extent to which axion couplings can be boosted in each mechanism, if at all, and explored their distinct predictions and phenomenology. In brief, $\mathcal{O}(10^{2}-10^3)$ coupling enhancements are possible by introducing large charges or vector kinetic mixing. Significantly larger enhancements are possible with clockwork models if one takes an agnostic view towards UV completions, but arbitrarily large amplifications may be stymied in string embeddings. Inversely, axion-axion kinetic mixing can only be effective if some string construction allows one to bypass the field theory arguments presented above.  Finally, discrete symmetries and axion-photon kinetic mixing are ineffective in raising the axion coupling to visible matter. If a discovery of ultralight axion dark matter is made by a search in the near future, it would be a clear sign of new dynamics  with possible implications for other low energy terrestrial experiments.  

There are several phenomena not discussed above that may place further restrictions on ultralight axion models. First of all, if a symmetry is restored in the early universe, topological defects such as domain walls and cosmic strings could form. Axion emission from cosmic strings would contribute to the energy density of axions present today and any stable domain walls may dominate the energy density and thereby drastically alter the cosmology. This may further constrain variants of axions models, such as clockwork axions, whose UV completions could have multiple restored symmetries. Additionally, if an axion symmetry is restored, axions may form miniclusters~\cite{Kolb:1993zz} which would contribute to dark matter small scale structure. These may be observed using probes such as microlensing~\cite{Fairbairn:2017sil}, pulsar timing~\cite{Dror:2019twh,Ramani:2020hdo}, and 21cm cosmology \cite{Kadota:2020ybe}. We leave the consideration of these issues for future work.

We considered ultralight axion dark matter, but the mechanisms discussed here may be applied in other contexts where large axion couplings to the visible sector are desirable. Some examples include inflation (where most of these mechanisms first arose, see text for references), looking for parametric resonance during axion minicluster mergers~\cite{Hertzberg:2018zte,Hertzberg:2020dbk}, monodromy axions~\cite{Jaeckel:2016qjp,Berges:2019dgr}, vector dark matter production~\cite{Agrawal:2018vin,Co:2018lka}, and addressing the $H_0$ tension~\cite{Gonzalez:2020fdy,Weiner:2020sxn}. Lastly, while we focused primarily on the axion-photon and axion-nucleon couplings, similar bounds can be constructed for axion-electron couplings and, potentially, ultralight neutrino-philic scalars~\cite{Berlin:2016woy} (whose potential likely also needs to arise from breaking of a shift symmetry to be protected against quantum corrections from gravity). We leave a study of such scalars to future work.

\section*{Acknowledgments}
We thank Tristan Smith for clarification of the bounds derived in~\cite{Poulin:2018dzj}, Anson Hook, Lawrence Hall, and Prateek Agrawal for helpful comments on the manuscript, and Nicholas Rodd, Gustavo Marques-Tavares, Katelin Schutz, and Matthew Moschella for useful discussions. This work was supported in part by the Director, Office of Science, Office of High Energy and Nuclear Physics, of the U.S. Department of Energy under Contract DE-AC02-05CH11231 (JD \& JML) and by the National Science Foundation under grant PHY-1316783 (JML)

 \bibliographystyle{JHEP}
\bibliography{draft}

\end{document}